\documentclass[fleqn,10pt,dvipsnames]{wlscirep}
\usepackage{hyperref}
\usepackage{graphicx,bm}
\usepackage{amssymb,amsmath}
\usepackage{xcolor}

\usepackage{physics}

\usepackage{newtxtext} %

\usepackage{subcaption}
\usepackage{float} %Used for H option in figures

\usepackage[colorinlistoftodos]{todonotes}

\title{Electron-Phonon Decoupling in Two Dimensions}

\author{George McArdle}
\author{Igor V. Lerner*}
\affil {University of Birmingham, School of Physics \& Astronomy, B15 2TT, UK}

\affil[*]{i.v.lerner@bham.ac.uk}

\begin{abstract}
 In order to observe many-body localisation in electronic systems, decoupling from the lattice phonons is required, which is possible only in out-of-equilibrium systems. We show that such an electron-phonon decoupling may happen in suspended films and it manifests itself via a bistability in the electron temperature.   By studying the electron-phonon cooling rate in disordered, suspended films with two-dimensional phonons, we derive the conditions needed for such a bistability,  which can be observed experimentally through hysteretic jumps of several orders of magnitude in the nonlinear current-voltage characteristics.  We demonstrate that such a regime is  achievable in systems with an Arrhenius form of the equilibrium conductivity, while practically unreachable in materials with Mott or Efros-Shklovskii hopping.
\end{abstract}
\begin{document}

\flushbottom
\maketitle

\thispagestyle{empty}

\section*{Introduction}
Tremendous experimental progress in isolating quantum many-body systems from the environment (see \cite{Bloch:08RMP} and \cite{Abanin:2019} for reviews) led to the observation of many-body localization (MBL) in ultracold atomic systems.\cite{Schreiber:2015,Schneider:17PRX} The question remains, however, whether MBL can be observed in disordered electronic systems for which it was originally predicted.\cite{BAA1,MGP:05} In the absence of interaction, disorder localizes all electron states in low-dimensional systems\cite{AALR} so that the dc electronic current vanishes without inelastic processes. The essence of MBL is that inelasticity due to {the electron-electron (e-e)} interaction alone does not lead to thermal equilibration of the system, as was first suggested for interacting electrons in a chaotic quantum dot.\cite{AlGKL}
 Hence in the absence of other mechanisms of inelasticity
 all states would remain localized so that finite-temperature conductivity would remain  zero.

The main  obstacle to the observation of this effect in electronic systems lies in the coupling of the electron system to the environment via the electron-phonon {(e-ph)} interaction. In equilibrium, such a coupling equilibrates all electron states with the underlying lattice leading to their delocalization. This results in nonzero finite-temperature conductivity, which is driven, in the absence of the electron-electron interaction, by Mott's variable-range hopping \cite{Mott:68,MottDavis} and given, at temperatures lower than some constant $T_0$, by
 \begin{align}\label{Mott}
    \sigma({T})=\sigma_0\exp\left[-({T_0/T})^{\gamma } \right],
 \end{align}
where $\gamma =1/(d+1)$ for a $d$-dimensional system, and $\sigma_0$ is a constant, temperature-independent prefactor. {The presence of an electron-electron interaction changes the mechanism of equilibration at sufficiently low temperatures} due to the emergence of a so-called Coulomb gap in the single-electron density of states\cite{Efr-Sh:75} resulting in the change of the exponent in Eq.~\eqref{Mott} to $\gamma = 1/2$, {independent of dimensionality}.

Although {the electron-phonon  coupling  makes it impossible to observe MBL in electronic systems equilibrated with the lattice, in out-of-equilibrium systems electrons and phonons might decouple even in the presence of a weak electron-phonon interaction. For MBL to be observable, the interacting electrons should be at internal equilibrium but not equilibrated with the underlying lattice.}  It has been suggested\cite{BAA2} that such an {out-of-equilibrium decoupling could manifest itself  via a bistability in the nonlinear current-voltage ($I$-$V$) characteristics.} It has been shown later\cite{AKLA:09} that such a bistability, caused by the electrons overheating, occurs at low temperatures, $T\lesssim0.1T_0$, provided that the equilibrium conductivity is close to  the Arrhenius law,   i.e.\ $\gamma \approx1$ in Eq.~\eqref{Mott}. The quantitative description of this bistability,\cite{AKLA:09}  based on an earlier developed analysis of the electron-phonon equilibration rate in bulk disordered systems,\cite{Reizer,Sergeev_Mitin,YK} allowed a full explanation of giant jumps (up to six orders in magnitude) of resistivity experimentally observed\cite{Shahar:05,Bat1,Bat2,Shahar:09} in various materials with the Arrhenius equilibrium conductivity where $T_0$ is of order of a few kelvins.

The presence of a bistability in the $I$-$V$ characteristics below a critical temperature is not, by itself, necessarily a signature of MBL but its absence would mean electron-phonon equilibration and hence the absence of MBL.  Intuitively, it seems that the electron-phonon decoupling would be easier to achieve in suspended disordered 2d films. Hence, such films might be promising for observing MBL provided that they are sufficiently disordered for the one-electron Anderson localization on the length scale smaller than the film dimensions.

In this paper, we derive the electron-phonon equilibration rate in such films and use it to analyze a possible bistability of the $I$-$V$ characteristics on the insulating side where the equilibrium conductivity is governed by Eq.~\eqref{Mott}. We found that in   suspended films with the Arrhenius equilibrium conductivity, the bistability occurs at lattice temperatures $T\lesssim0.1T_0$   similar, albeit quantitatively different, to bulk systems or thick multilayered films. On the other hand, for disordered films with $\gamma \lesssim1/2$, i.e.\ those with either Mott\cite{Mott:68,MottDavis} or Efros-Shklovskii\cite{Efr-Sh:75} conductivity, the bistability could take place at much lower temperatures. Hence   only materials with the  Arrhenius resistivity at low temperatures could be potentially promising for detecting MBL. While the origin of a small Arrhenius gap, $T_0\sim1$K, is quite an interesting problem by itself, we do not consider it here noticing only that there is a variety of materials with such a gap \cite{KowOva:94,Gantmakher:96,Shahar:04,AvishaiMeir:07,Palevski:21}  which typically have granular disorder.

\section*{Model}
We consider electron-phonon relaxation in a suspended disordered film where both electron and phonon degrees of freedom are two-dimensional. Electrons can thermally decouple from phonons when a finite source-drain voltage, $V$, drives the system out of equilibrium. The decoupling might reveal itself in a nonlinear, non-Ohmic regime when  the electron-phonon interaction is too weak to effectively dissipate the power supplied to the electron system.
 Assuming  the electron-electron interaction to be sufficiently strong for thermalizing electrons between themselves at a temperature $T_{\mathrm{el}}$, the  energy dissipation from the electronic system to the phonon bath (or equivalently the lattice), which is at a temperature $T_{\mathrm{ph}}$, can be described  by the phenomenological heat balance equation,\cite{AKLA:09}
\begin{align}\label{HB}
 \frac{V^{2}}{R(T_{\mathrm{el}})}= \dv{\mathcal{E}}{t} =
 \frac{\mathcal{E}({T_{\mathrm{el}}})} {\tau_{\mathrm{e}\text-{\mathrm{ph}}}({T_{\mathrm{el}}})}
 -\frac{\mathcal{E}({T_{\mathrm{ph}}})} {\tau_{\mathrm{e}\text-{\mathrm{ph}}}({T_{\mathrm{ph}}})}.
\end{align}
Here the temperature-dependent part of the total electron energy is given by $\mathcal{E}{(T )} = \pi^2 \nu   \mathcal{A}T ^2/6$ (where ${\mathcal{A}}$ is the sample area and $\nu$ is the density of states at the Fermi surface), and $R({T_{\mathrm{el}}})$ is the sample resistance at equilibrium,  which is equal to the inverse conductivity $\sigma^{-1}({T_{\mathrm{el}}}) $, see Eq.~\eqref{Mott},  assuming for simplicity a square shape of the film.
As the electron  energy is conserved in  e-e collisions, the heat balance is fully determined by the e-ph interaction with the scattering time $\tau_{\mathrm{e}\text-{\mathrm{ph}}}(T)$ which is energy-independent at the low temperatures at which MBL might occur, as the relevant part of the dispersion for both the electrons and phonons is linear.  In the presence of disorder, the e-ph interaction is modified by the effect of phonon-induced impurity displacements.\cite{Reizer,Sergeev_Mitin,YK,Schmid} This can occur in two possible ways depending on whether the phonons directly affect the impurities. In the case of a suspended film, the impurities oscillate with the lattice so that the Hamiltonian becomes
\begin{align}\label{e-ph}
    {\mathcal{H}}&=\frac{1}{\sqrt{\mathcal{A}}}\sum_{{\bm{p}}, {\bm{q}}, {\bm{k}}} c^\dagger _{\bm{p}+\bm q+\bm k}c^{{\phantom{\dagger }}} _{\bm{p}} \big({\bm{g_q}} \cdot {\bm{u_q}}\,\delta _{\bm{k0}} + {\bm{g}}^{{\mathrm{imp}}}_{\bm{k}} \cdot {\bm{u_q}}  \big).
\end{align}
Here $c^\dagger , \ c $  are the electron  creation and annihilation operators, ${\bm{u_q}}$ is the the Fourier transform of the lattice displacement (corresponding to either transverse or longitudinal phonons), ${\bm{g_q}}=iC{\bm{q}}$ is the standard electron-phonon vertex with the deformation potential $C$ equal to the Fermi energy $\varepsilon _{\mathrm{F}}$ for two-dimensional phonons, and $ {\bm{g}}^{{\mathrm{imp}}}_{\bm{k}}=-iU({{\bm{k}}}){\bm{k}}$ is the vertex corresponding to {the phonon-displaced} impurities, with $U({{\bm{k}}})$ being the Fourier transform of the impurity potential. For electron scattering from impurities   we assume the standard model  of uncorrelated $s$-scatterers,\cite{AGD} which is equivalent to the Gaussian potential with zero average and  $\delta $-correlations,
\begin{align}\label{dis}
 \left< U({\bm r})U({\bm r'})\right>=\frac1{2\pi \nu \tau}  \delta ({\bm r-\bm r'}),
\end{align}
where $\tau$ is the mean scattering time.

 \section*{Results}
We show that {electrons can decouple from the phonon bath in thin suspended films provided that the equilibrium finite-temperature conductivity is close to the Arrhenius law, i.e.\ $\gamma \approx1$ in Eq.~\eqref{Mott}, and the bath temperature is much lower than the Arrhenius ``gap'' $T_0$}. This conclusion is based on our analysis of the electron-phonon cooling rate for 2d phonons similar to that for the phonons in bulk materials (see, e.g., \cite{Reizer, Sergeev_Mitin,YK}). Using the quantum kinetic equation derived in the Keldysh formalism (see, e.g.,  \cite{RS:86}), we derive the following expression for the electron-phonon cooling rate due to transverse phonons:
\begin{align}\label{e-ph-cooling}
	 \dot{\mathcal{E} } =  \frac{\alpha^2 k_{\mathrm{F}} \ell  n_{\mathrm{el}}\mathcal{A}}{\hbar\Delta_0^3}\left(T^5_{\mathrm{el}}-T^5_{\mathrm{ph}}\right), \hspace{10pt} \Delta_0^3 = \hbar^2 \rho_{2d} u_{\mathrm{t}}^4, \hspace{10pt} \alpha^2 = \frac{3}{\pi}\zeta(5)\approx0.99,
\end{align}
where $k_{\mathrm{F}}$ is the Fermi wave vector, $\ell $ is the electron mean free path, $u_{\mathrm{t}}$ is the transverse phonon speed of sound, $n_{\mathrm{el}}=k_{\mathrm{F}}^2/(2\pi)$ is the 2d electron density, $\rho_{2d}$ is the 2d material density and $\zeta$ is the Riemann-zeta function. This result corresponds to the $\tau$-approximation for the e-ph relaxation rate in Eq.~\eqref{HB} with the temperature dependence   $1/ {\tau_{\mathrm{e}\text-{\mathrm{ph}}}}(T)\propto T^3$ and is similar to that for the case of 3d phonons,\cite{AKLA:09}   where $\dot{\mathcal{E}} \propto T_{\mathrm{el}}^6 - T_{\mathrm{ph}}^6$, with the difference being caused by the weaker dependence of the phonon density of states on the phonon frequency, which goes as $\omega^{d-1}$.  As in the  3d  case, the leading contribution to the  cooling rate is due to the impurity-facilitated interaction of electrons with transverse phonons, which is absent in a clean metal. The contribution from the interaction with  longitudinal phonons has the same form as Eq.~\eqref{e-ph-cooling} with the change $u_{\mathrm{t}} \rightarrow u_{\mathrm{l}}$. Since the longitudinal speed of sound, $u_{\mathrm{l}}$, is typically a few times larger than its transverse counterpart, \cite{Gershenson} the longitudinal-phonon contribution contains a small factor of $ (u_{\mathrm{{t }}}/u_{\mathrm{{ l}}})^4$ in comparison to  the leading contribution given by Eq.~\eqref{e-ph-cooling}.  It is worth noting that the overall low-temperature suppression of the e-ph relaxation rate in disordered semiconductors, as compared to a clean metal,  is given by a factor of $n^*T\ell /\hbar u_{\mathrm{t}} (u_{\mathrm{{l }}}/u_{\mathrm{{ t}}})^3$, reflecting Pippard's ineffectiveness condition. \cite{Kittel_Book} Here $n^*$ is the number of electrons per unit cell, which is small in semiconductors most promising for MBL so that, with a typical $u_{\mathrm{t}}$ of order of $10^3{\mathrm{m/s}}$, the cooling rate could be several orders in magnitude smaller than in a dense clean metal in spite of the factor $(u_{\mathrm{{l }}}/u_{\mathrm{{ t}}})^3\sim10$.

 Next, we  substitute the cooling rate  \eqref{e-ph-cooling} into the heat balance equation \eqref{HB}. Assuming  the usual Drude prefactor for the equilibrium resistance,
\begin{align}\label{Drude}
R(T_{\mathrm{el}})=R_0 \exp[{ \left({\frac{T_0} {T_{\mathrm{el}}}}\right)^{\!\!\gamma } }]\equiv \frac{\hbar k_{\mathrm{F}}}{n_{\mathrm{el}}e^2 \ell}\exp[{ \left({\frac{T_0} {T_{\mathrm{el}}}}\right)^{\!\!\gamma } }],
\end{align}
we find that the heat balance equation is independent of the mean free path, $\ell$. This allows us to extend the results for  the electron-phonon cooling rate we have obtained in the metallic regime, $k_{\mathrm{F}}\ell\gg1$,  to the transition regime, $k_{\mathrm{F}}\ell\sim1$, and beyond. This is empirically justified by experiments\cite{Shahar:09} made in the vicinity of the superconducting-insulating transition, where $k_{\mathrm{F}}\ell<1$, as the results obtained were in excellent quantitative agreement with the results for the bistability\cite{AKLA:09} obtained using the cooling rate via interactions with bulk  phonons which had been calculated in the metallic regime.\cite{Reizer,Sergeev_Mitin,YK}

 It is convenient to represent the heat balance equation, obtained by substituting the equilibrium resistance \eqref{Drude} and  the cooling rate  \eqref{e-ph-cooling} into Eq.~\eqref{HB}, in terms of a dimensionless temperature and voltage, defined by $t_{{\mathrm{el,ph}}} = T_{\mathrm{el,ph}}/T_0$ and $v = V/V_0$ with $V_0^2 = \alpha^2 k_{\mathrm{F}}^2\mathcal{A} T_0^5 /(e^2 \Delta_0^3)$, as follows:
\begin{align}\label{HB:dimensionless}
	v^2= \left[t_{\mathrm{el}}^5 - t_{\mathrm{ph}}^5\right]\exp\left[(1/t_{\mathrm{el}})^\gamma\right] .
\end{align}
For any given voltage, the electron temperature must be higher than the bath temperature to satisfy this equation. By itself this does not signify the electron-phonon decoupling. On the other hand, we can see clear evidence of decoupling in the presence of a bistability where, below a critical bath temperature and  in a certain range of the applied voltage, electrons can mutually equilibrate  at two distinct temperatures, ``cold'' $t_{\mathrm{el}}^<$ and  ``hot'' $t_{\mathrm{el}}^>$. It is in the regime of overheating, at temperature $t_{\mathrm{el}}^>$ which is practically independent of the lattice temperature $t_{{\mathrm{ph}}} $, that the electrons become fully decoupled from the phonons.

\begin{figure}[ht]
	\centering
	\includegraphics[width = 0.45 \textwidth]{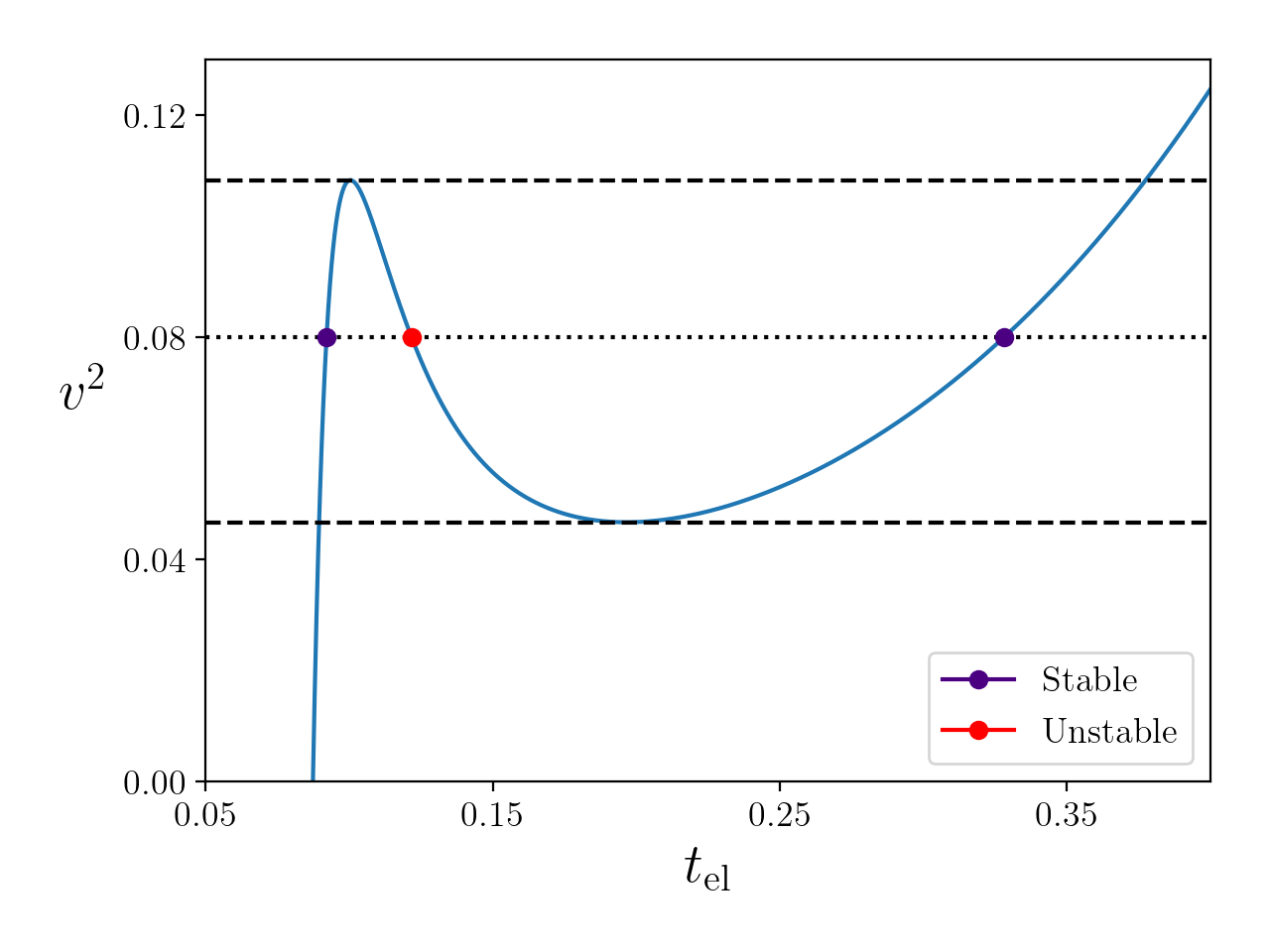}\qquad\includegraphics[width = 0.45 \textwidth]{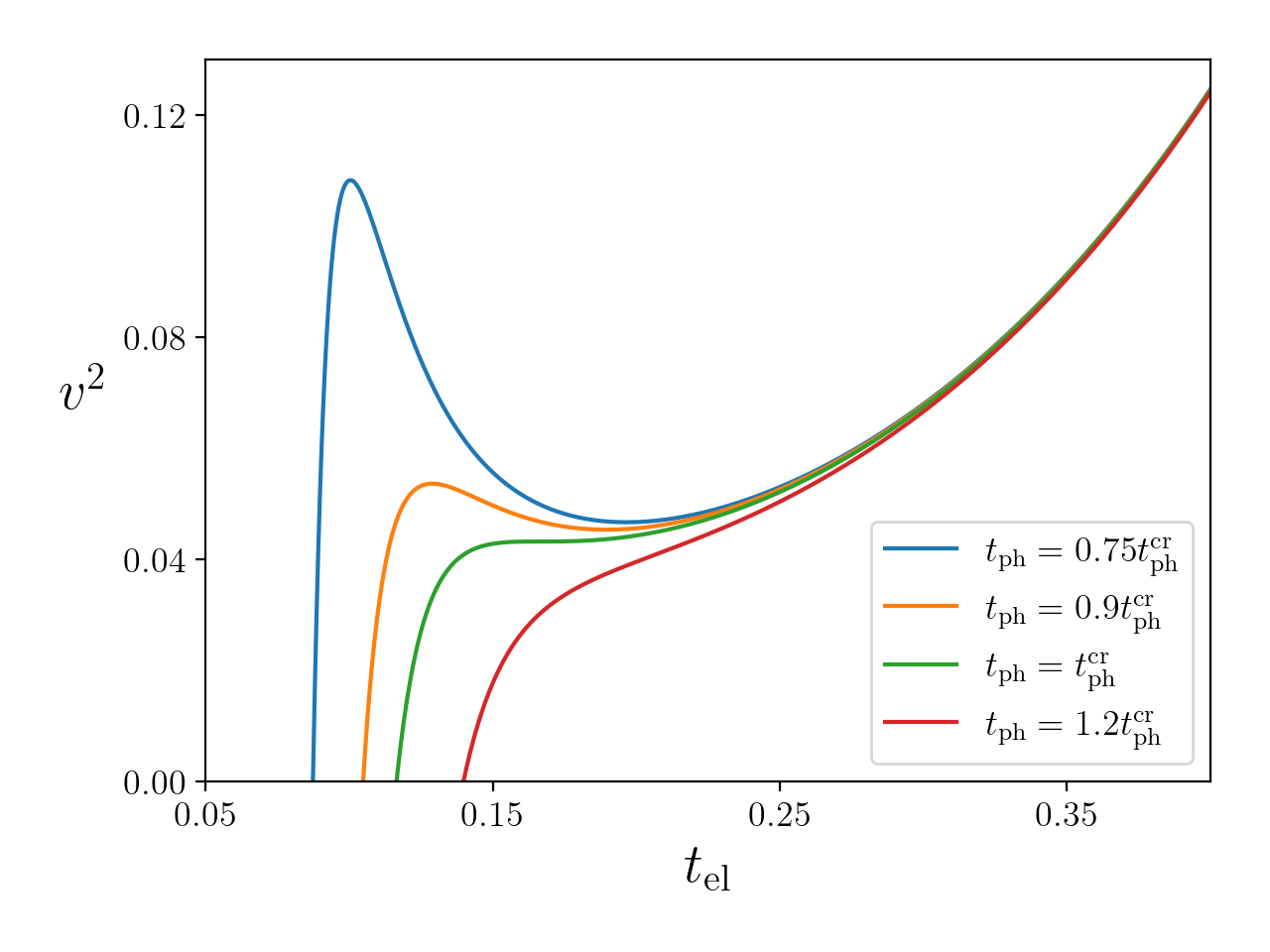}
\textbf{({a})}\hspace*{.5\textwidth}\textbf{({b})}
	\caption{\label{fig:Bistability}\textbf{(a).} The bistability region, where two stable solutions for $t_{\mathrm{el}}$ exist  in a certain range of the applied voltage, is shown for $t_{\mathrm{ph}} =0.75 t_{\mathrm{ph}}^{\mathrm{cr}}$ for the Arrhenius equilibrium resistance, $\gamma =1$. The blue dots correspond to cold and hot electron states at   temperatures $t_{\mathrm{el}}^{< }$ and $t_{\mathrm{el}}^{> }$, respectively, and the red dot to an unstable solution. \textbf{(b).} The dependence of $t_{\mathrm{el}}$ on $v^2$ for different phonon bath temperatures is shown as follows from Eq.~\eqref{HB:dimensionless}. Above the critical bath temperature this corresponds to the actual $t_{\mathrm{el}}({v})$ dependence while below $ t_{\mathrm{ph}}^{\mathrm{cr}}$ the electronic system will fall either to  {$t_{\mathrm{el}}^<$ or to $t_{\mathrm{el}}^>$,} making temperatures in between experimentally inaccessible.}
\end{figure}

 {Such a  bistability occurs when Eq.~\eqref{HB:dimensionless} has,  for a given voltage and bath temperature, two stable solutions for $t_{\mathrm{el}}$. This happens below the critical phonon bath temperature $t_{\mathrm{ph}}^{\mathrm{cr}} $ when the r.h.s.\ of this equation becomes a non-monotonic function of the electron temperature. An elementary analysis shows that the critical bath temperature is  given by
\begin{align}\label{t_crit}
	t_{\mathrm{ph}}^{\mathrm{cr}} \equiv \frac{T_{\mathrm{ph}}^{\mathrm{cr}}}{T_0}= \left(1+\frac{5}{\gamma}\right)^{-\left(\frac1\gamma+\frac15\right) }.
\end{align}
For $t_{\mathrm{ph}}<t_{\mathrm{ph}}^{\mathrm{cr}}$ Eq.~\eqref{HB:dimensionless} has three solutions in a certain region of the source-drain voltage, as illustrated in Fig.~\ref{fig:Bistability} for the Arrhenius case, $\gamma =1$. For a given voltage within this region, both the ``cold'' and ``hot'' states, at electronic temperatures $t_{\mathrm{el}}^<$ and $t_{\mathrm{el}}^>$ respectively, are stable. The middle solution, however, corresponds to an unstable electronic state.

Formally, a similar bistability takes place also for the Mott  ($\gamma = 1/3$) and Efros-Shklovskii ($\gamma = 1/2$) hopping regimes. However, a faster than exponential dependence of the critical phonon bath temperature on $1/\gamma $, Eq.~\eqref{t_crit}, pushes the bistability in these regimes to very low temperatures: while $t_{\mathrm{ph}}^{\mathrm{cr}} \approx 0.12$ in the Arrhenius case, it is about $ 5\cdot 10^{-3} $ in the Efros-Shklovskii regime, and $10^{-4} $ in the Mott regime. With $T_0\sim 1$K
 in materials of interest, the bistability regime would be practically  unreachable  in the systems with the Mott or Efros-Shklovskii conductivity, while the experimentally observed bistability in an Arrhenius material \cite{Shahar:09} was in a quantitative agreement with the theoretical description \cite{AKLA:09} similar to that developed here but with the electrons interacting with bulk phonons.  Due to this fact we conclude that  for $\gamma \approx 1$ the bistability occurs for $T \lesssim 0.1T_0$.

 At the bistability boundaries for a given $t_{\mathrm{ph}}$, the derivative of the r.h.s.\ of Eq.~\eqref{HB:dimensionless} vanishes, so that the boundaries are determined in the Arrhenius case by the following equation
 \begin{align}\label{BSB}
    5t_{\mathrm{el}}=1-({t_{\mathrm{ph}}/t_{\mathrm{el}}})^5,
 \end{align}
which for $t_{\mathrm{ph}}<t_{\mathrm{ph}}^{\mathrm{cr}}$ has two solutions,  hot, $t_{\mathrm{el}}^{\mathrm{h}}$, and cold, $t_{\mathrm{el}}^{\mathrm{c}}$, depicted in Fig.~\ref{fig:Boundaries}(a). The corresponding temperature dependence of the voltage boundaries of the bistability, $v^>$ for the cold state and $v^<$ for the hot one, obtained by substituting $t_{\mathrm{el}}^{{\mathrm{c,h}}} $ into Eq.~\eqref{HB:dimensionless}, is shown  in Fig.~\ref{fig:Boundaries}(b).  As previously mentioned, in order to satisfy the heat balance, Eq.~\eqref{HB}, the electron temperature must always be higher than the phonon bath. However, while in the cold state $t_{\mathrm{el}}^c$ almost follows $t_{\mathrm{ph}}$, in the overheated hot state $t_{\mathrm{el}}^h$ is almost independent of the bath temperature, and so is the voltage boundary of this state, $v^<$. Since the electrons in the overheated state are practically decoupled from the phonon bath, it is the state most suitable for a possible observation of MBL.
 \begin{figure}[ht]
	\centering
	\includegraphics[width = 0.45 \textwidth]{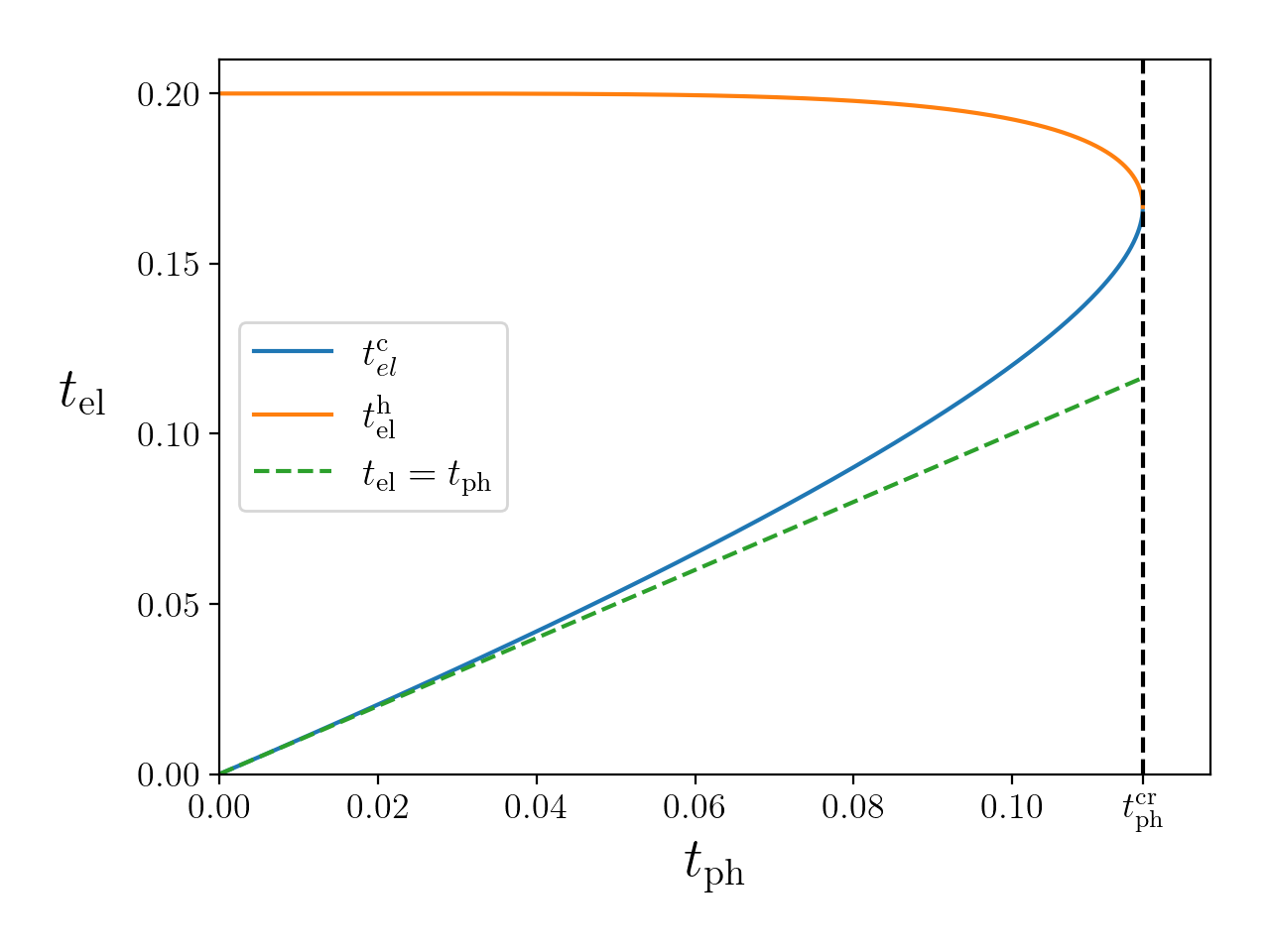}\qquad\includegraphics[width = 0.45 \textwidth]{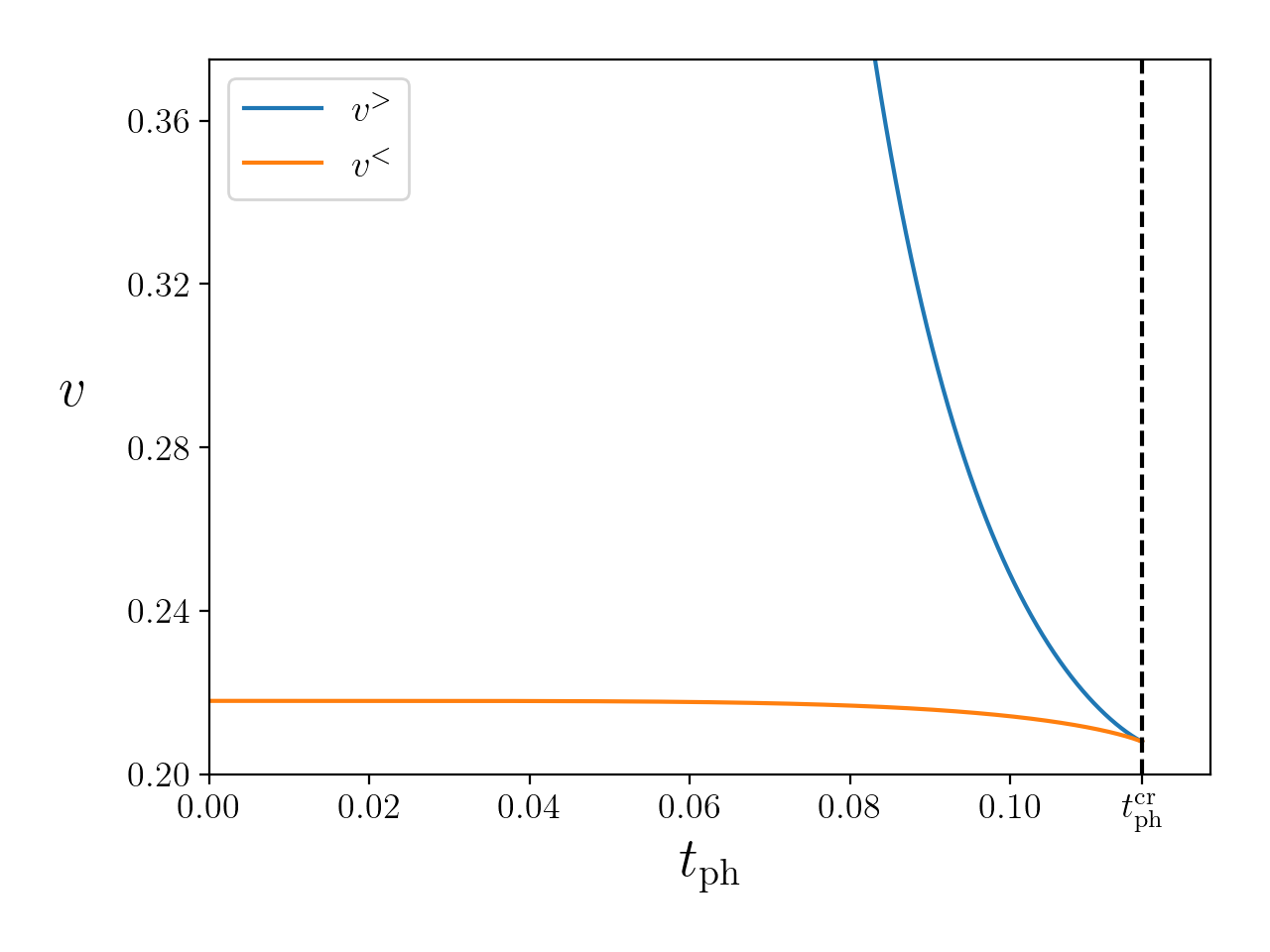}
	\textbf{({a})}\hspace*{.5\textwidth}\textbf{({b})}
\caption{\label{fig:Boundaries} Dependence of the bistability boundaries on the phonon temperature for \textbf{({a})} the electron temperature and \textbf{({b})} the source-drain voltage, for $\gamma=1$. The region of electron temperatures inside the curve \textbf{({a})} is experimentally inaccessible as it corresponds to the unstable states.}
\end{figure}

It is important to understand the experimental signatures of the bistability as this will confirm with certainty that electron-phonon decoupling is present.
A relatively simple experimental signature is the presence of a region of excluded temperatures corresponding to the unstable states,  which are those enclosed by the curve in Fig.~\ref{fig:Boundaries}(a). Such a region was experimentally observed in \cite{Shahar:09} and turned out to be in quantitative agreement with the theoretical prediction \cite{AKLA:09} made for films on a substrate with electrons interacting with bulk phonons. But the most striking feature due to bistability is giant hysteretic jumps in the $I$-$V$ characteristics: due to the exponential dependence of  resistance on the inverse electron temperature, a switch between the cold and hot electron states under a given voltage $V$ leads to abrupt changes in the current $I$ that can be of many orders in magnitude.

To see this, we solve numerically  the equation for the non-linear conductance  {in the Arrhenius regime,}
\begin{align}\label{conductance}
	G = \frac{I}{V} = \frac{1}{R(t_{\mathrm{el}})} = \frac{1}{R_0}\mathrm{e}^{-1/t_{\mathrm{el}}(v)},
\end{align}
where $R_0$ is the Drude resistivity, Eq.~\eqref{Drude}.   The solution has an $S$-shape, as shown in Fig.~\ref{fig:IV}(a), with the dotted part being unstable. This makes hysteretic jumps between the low conductance (cold electron) state and the high conductance (hot electron) state inevitable.

These jumps are illustrated in  Fig.~\ref{fig:IV}(b). Let us stress that exact positions of the jumps are random as
 the boundaries here are simply bounds on the true jumps; where the actual jumps occur depends on the decay mechanisms of the states, as discussed  in \cite{AKLA:09, Doron_2017}.  Moreover, we  do not estimate numerical values for these boundaries, because in order to obtain an accurate value for the voltage scale, $V_0$, we would also need to include the effects of localisation into the electron-phonon cooling rate\cite{KravFeigel}, which goes beyond the aim of this work. Despite this, the temperature dependence of the positions of the jumps should be   experimentally observable, as in the case of electrons interacting with bulk phonons. \cite{Shahar:05,Bat1,Bat2,Shahar:09}  Namely, one expects to see a strong temperature dependence of the boundary for the cold electron states $\left(v^>\right)$ and almost no temperature dependence of the boundary for the hot states $\left(v^<\right)$, as well as the inaccessible region of electron temperatures as in Fig.~\ref{fig:Boundaries}(a).

\begin{figure}[ht]
	\centering
	\includegraphics[width = 0.45 \textwidth]{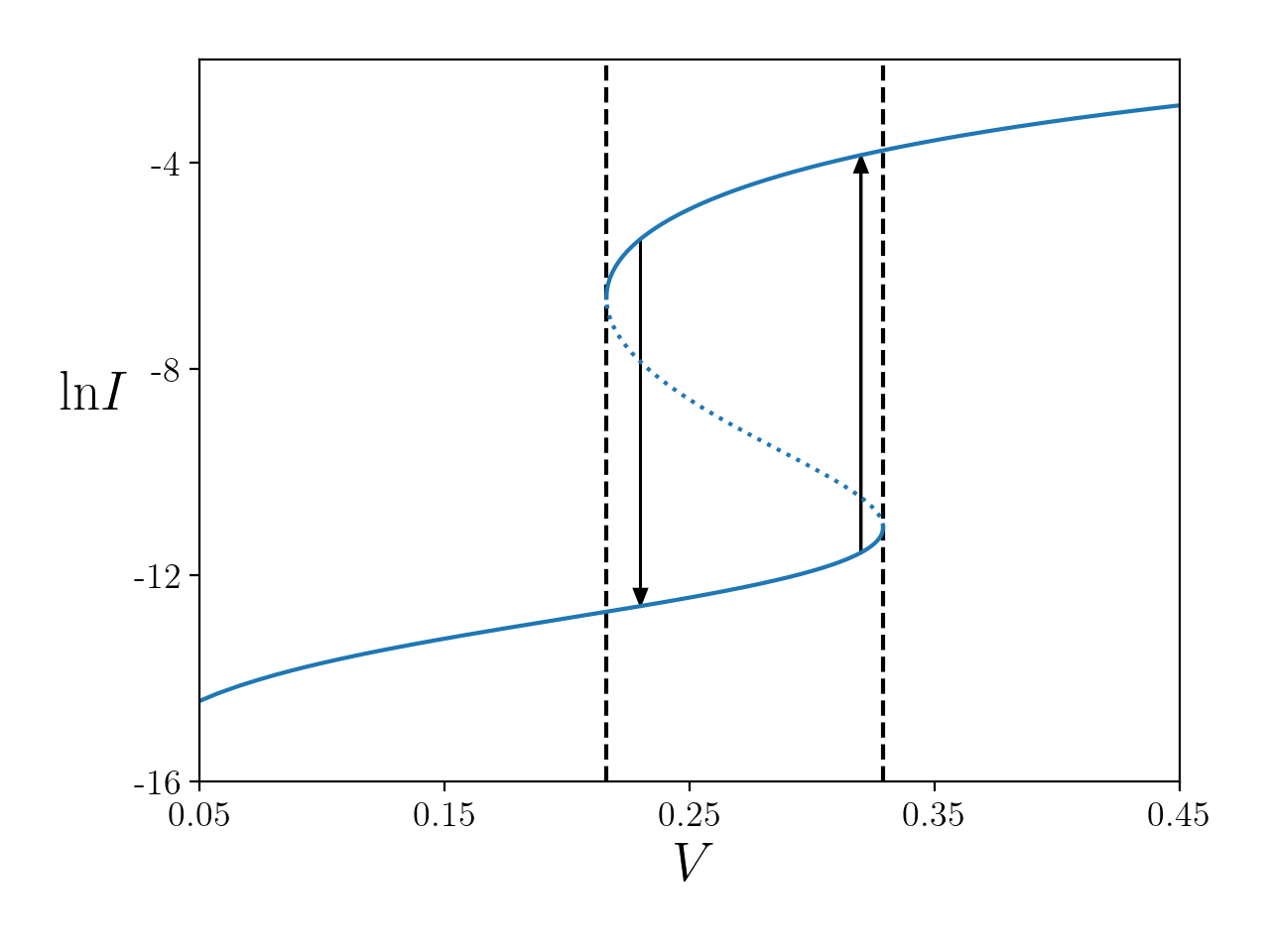}\qquad\includegraphics[width = 0.45 \textwidth]{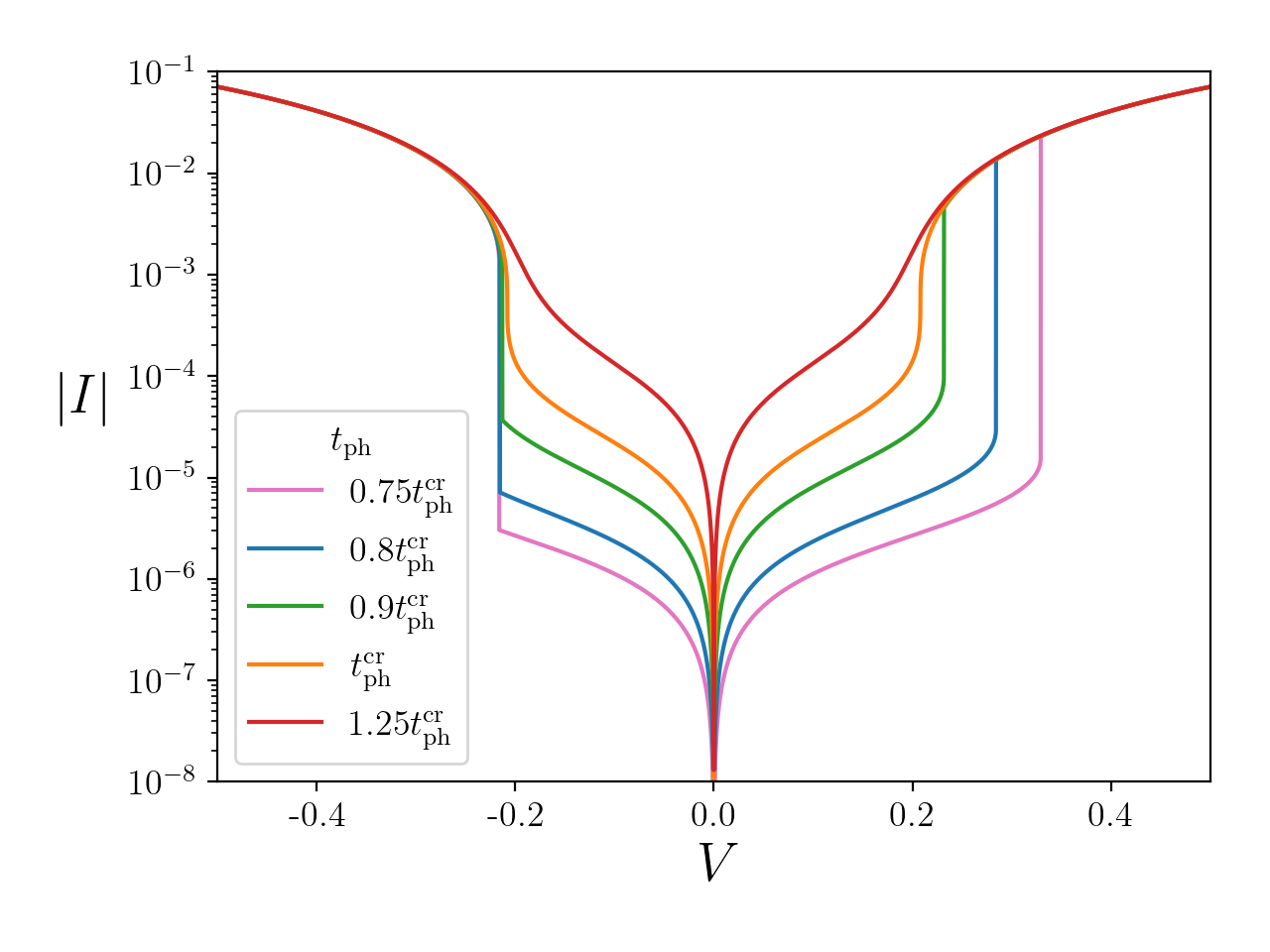}
	\textbf{({a})}\hspace*{.5\textwidth}\textbf{({b})}
	\caption{\label{fig:IV}{\textbf{(a).} The $S$-shape solution to the equation for the non-linear conductance, Eq.~\eqref{conductance}, for $t_{\mathrm{ph}} =0.75 t_{\mathrm{ph}}^{\mathrm{cr}}$. The dotted part corresponds to unstable states, resulting in hysteretic jumps, denoted by the arrows. Note that the jumps do not necessarily occur at the boundaries of the bistability (dashed lines).  \textbf{(b).} The numerically predicted $I$-$V$ characteristics for various lattice temperatures. The jumps here are shown to be at the bistability boundaries, though this may not be the case in reality. The $V>0$ side of the graph illustrates the transition from the cold electron (low conductance) state to the hot electron (high conductance) state, which occurs when the source-drain voltage is increased. The $V<0$ side displays the opposite transition when the voltage is decreased, going from the hot to cold electron states. In both \textbf{(a)} and \textbf{(b)} the voltage is measured in units of $V_0$ and the current is in units such that the resistance is measured in units of $R_0$. }}
\end{figure}

\section*{Discussion}

To summarise, we have shown that for films with an equilibrium conductivity exhibiting an Arrhenius (or Arrhenius-like) law, $R(T)=R_0\exp\left[({T_0/T})^{\gamma } \right]$ with $\gamma \approx 1$, electrons can decouple from phonons in a nonlinear regime. Such an electron-phonon decoupling manifests itself as a bistability in the electron temperature that can be observed  via   the $I$-$V$ characteristics. This bistability occurs  in a certain region of source-drain voltages for a lattice temperature $T \lesssim 0.1T_0$, while  $T_0$ is known to be of order of a few kelvins in numerous materials. On the contrary, in systems where the equilibrium conductivity is of the form of either Mott $\left(\gamma=1/3\right)$ or Efros-Shklovskii $\left(\gamma=1/2\right)$ hopping, the lattice temperature required for the bistability is much lower and practically not accessible. Therefore, for achieving  the electron-phonon decoupling necessary for MBL, materials with an Arrhenius conductivity, such as those recently seen in \cite{Gul_Holmes}, are most promising.

The bistability in the electron temperature means that there exist  stable `cold' and `hot' electron states. The former have a  temperature proportional to (but slightly higher than) that of the lattice while the latter have a temperature which is almost independent of the lattice temperature. It is in this  state that the electrons are fully decoupled from the phonons, making it most promising  for  observing MBL. The most significant experimental signature of the bistability is giant jumps in the non-linear $I$-$V$ characteristics between the cold (low conductance) and the hot (relatively high conductance) states. Such jumps have been previously associated,  in the 3d case, with a possible transition to MBL\cite{Ovadia:15}. We emphasise, however, that while these jumps provide the evidence for electron-phonon decoupling, further evidence would be needed to confirm the existence of the MBL state.

\section*{Methods}

In order to calculate the electron-phonon cooling rate in two-dimensional systems, Eq.~\eqref{e-ph-cooling}, we used the Keldysh formalism (see, e.g.,  \cite{RS:86}) in the form similar to that used in calculating the cooling rate in 3D systems.\cite{YK} The quantum kinetic equation can be written as
\begin{align}\label{Kin_eqn}
	\partial_{t} f_\varepsilon (t) = I[f],
\end{align}
where (after setting $\hbar=1$) the collision integral for the electron-phonon interaction modified by disorder is given by, \cite{YK}
\begin{align}\label{collision}
	I[f] = \frac{i}{4\pi\nu\mathcal{A}} \biggl<\int_{-\infty}^{\infty} \frac{{\mathrm{d}} \omega}{2\pi}\int {\mathrm{d}}\bm{r} {\mathrm{d}}\bm{r'} \Delta G(\bm{r}, \bm{r'},\varepsilon ) \hat{g}_\alpha (\bm{r'}) \Delta G(\bm{r'}, \bm{r},\varepsilon -\omega) \Delta D_{\alpha \beta}(\bm{r'}-\bm{r},\omega) \hat{g}_\beta (\bm{r})\\
	\times [(f_\varepsilon -f_{\varepsilon -\omega})N_\omega  +f_\varepsilon  f_{\varepsilon -\omega}-1]\biggr>.\notag
\end{align}
Here the brackets $\left<\cdots \right>$ stand for averaging over the disorder potential, Eq.~\eqref{dis}, $N_\omega = 1+2n_B(\omega)$ and $f_\varepsilon =1-2n_F(\varepsilon )$, with $n_B({\omega })$ and $n_F({\varepsilon })$ being the standard Bose and Fermi distributions respectively; $\hat{g}_{\alpha,\beta}$ can be either $\bm{g}_{\bm{q}}$ or $\bm{g}_{\bm{k}}^{\mathrm{imp}}$, see Eq.~\eqref{e-ph}; $\Delta G\equiv G^R-G^A$ and $\Delta D\equiv D^R-D^A$ are the differences between the retarded and advanced Green's functions for electrons and phonons, respectively.

The phonon Green's functions are not directly affected by impurities so that their Fourier transforms, which include contributions from the longitudinal, $j=\mathrm{l}$,  and transverse, $j=\mathrm{t}$,  phonons,  $\Delta D_{\alpha \beta}(\bm{q},\omega) = \sum_j \Delta D_{\alpha \beta}^{(j)}(\bm{q}, \omega)$, are given by the standard expressions
\begin{align}
	\Delta D_{\alpha \beta}^{(j)}(\bm{q}, \omega) = \left[D^R_{\alpha \beta}(\bm{q}, \omega)-D^A_{\alpha \beta}(\bm{q}, \omega) \right]^{({j})} = -\frac{\pi i\eta_{\alpha \beta}^{(j)}}{\rho_{2d} \omega_j(q)}\left[\delta\left(\omega - \omega_j(\bm q)\right) - \delta\left(\omega + \omega_j(\bm q) \right) \right] ,
\end{align}
 where  $\eta_{\alpha \beta}^{(l)} = q_\alpha q_\beta/q^2$  and $\eta_{\alpha \beta}^{(t)} = \delta_{\alpha \beta} - q_\alpha q_\beta/q^2$, and we assume the Debye model for the phonon dispersion, $\omega_j(\bm{q})=u_j|\bm q|\Theta({q_0-|\bm q|})$, where $q_0$ is the Debye momentum.

  The disorder-averaged electron Green's functions $G^{R,A}({\bm r, \bm r' , \varepsilon }) $ depend only on the difference of their spatial arguments, and the appropriate Fourier transforms are  given by
\begin{align}
	G^{R,A}(\bm{p}, \varepsilon ) = \frac{1}{\varepsilon -\xi_{\bm{p}}\pm i/2\tau}, \hspace{10pt} \xi_{\bm{p}} = \varepsilon_{\bm{p}} - \varepsilon_{\mathrm{F}}.
\end{align}
A further contribution of disorder in the collision integral \eqref{collision} is described by  vertex corrections. Including only the leading transverse phonons  contribution, these corrections are shown  in the metallic regime, ${k_{\mathrm{F}}}\ell \gg1$,  in Fig.\ref{fig:Diagrams}. In the absence of disorder, transverse phonons do not alter the local charge density and so cannot couple directly to the electrons. However, in disordered materials they contribute via the vertices $ {\bm{g}}^{{\mathrm{imp}}}_{\bm{k}}$, Eq.~\eqref{e-ph}, which describe the effect of phonon-induced impurity displacements.
\begin{figure}[ht]
	\centering
	\begin{subfigure}{0.5 \textwidth}
		\centering
		\includegraphics[width = 0.7 \textwidth]{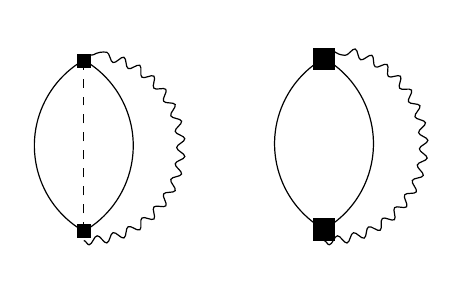}
	\end{subfigure}%
	\begin{subfigure}{0.5 \textwidth}
		\centering
		\includegraphics[width = 1 \textwidth]{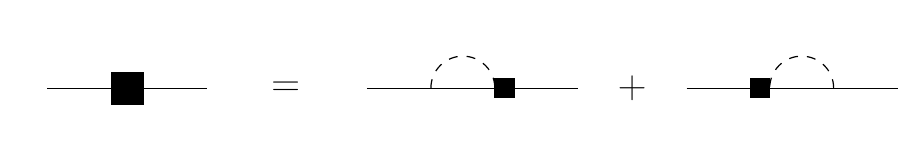}
	\end{subfigure}
	
	\caption{\label{fig:Diagrams} The two lowest-order diagrams that contribute to the collision integral in Eq.~\eqref{collision} due to the interaction of electrons with transverse phonons via impurity scattering: the smaller squares correspond to  $\bm{g}_{\bm{k}}^{\mathrm{imp}}$, the straight lines are the electron Green's functions, the wavy lines are the phonon Green's functions and the dashed lines describe the standard  averaging over impurities.}
\end{figure}

The longitudinal-phonons contribution to the cooling rate turns out to be functionally the same as that of the transverse phonons, given in Eq.~\eqref{e-ph-cooling},   with $u_{\mathrm{l}}$ substituted for $u_{\mathrm{t}}$. It is much smaller as $\left(u_{\mathrm{t}}/u_{\mathrm{l}}\right)^4\ll1$. Note that this contribution, which  exists also in clean systems,  involves  more cumbersome diagrams that include diffuson propagators  similar to the  3d  case.\cite{YK}
We do not give any further detail of calculating the longitudinal-phonons contribution as it is not relevant for the final results.

The calculation of the diagrams depicted in Fig.~\ref{fig:Diagrams} is relatively straightforward. We assume that the electron-electron interaction is sufficiently strong such that the electrons mutually thermalise and can be assigned a single temperature, $T_{\mathrm{el}}$ which is higher than the phonon bath (lattice) temperature, $T_{\mathrm{ph}}$. This results in a quasi-equilibrium situation where $f_\varepsilon  = \tanh(\varepsilon /2T_{\mathrm{el}}(t))$ and $N_\omega(T_{\mathrm{ph}}) = \coth (\omega/2T_{\mathrm{ph}})$. Then the spatial integral in Eq.~\eqref{collision} is calculated after the Fourier transform and using the fact that $q_T\ell \ll1$ (where $q_T\sim T/u_{\mathrm{t}}$ is a typical phonon momentum at temperature $T$) and the identity $f_\varepsilon  f_{\varepsilon -\omega} - 1=-N_\omega(T_{\mathrm{el}})(f_\varepsilon  - f_{\varepsilon -\omega})$, one reduces the collision integral to
\begin{align}\label{collision_integral_K}
	I[f] = \int \!{\mathrm{d}}\omega\, K(\omega) \left[N_\omega(T_{\mathrm{ph}}) - N_\omega(T_{\mathrm{el}}) \right] \left[(f_{\varepsilon +\omega} -f_\varepsilon ) + (f_{\varepsilon -\omega} - f_\varepsilon  )\right],
\end{align}
where $K(\omega)$ is expressed in terms of a dimensionless electron-phonon coupling constant, $\beta_{t}=  \dfrac{\nu\varepsilon_{\mathrm{F}}^2}{2\rho_{2d} u_{\mathrm{t}}^2},$ as
\begin{align}\label{transverse}
	K(\omega) = \frac{\beta_{t}\operatorname{sgn}(\omega) }{8k_{\mathrm{F}} \ell} \left( \frac{\omega \ell}{u_{\mathrm{t}}} \right)^2.
\end{align}
 Substituting the result of Eq.~\eqref{collision_integral_K} into Eq.~\eqref{Kin_eqn} and multiplying both sides by $\varepsilon $, one finds after integrating with to respect to $\varepsilon $ that the cooling rate (restoring factors of $\hbar$) is given by
\begin{align}\label{Eq:Cooling integral}
	\dot{\mathcal{E}} = \frac{k_{\mathrm{F}} \ell n_{\mathrm{el}}\mathcal{A}}{\hbar \Delta_0^3}\int_0^\infty \frac{d\omega}{16\pi} \omega^4 \left[ \coth \left( \frac{\omega}{2T_{\mathrm{el}}}\right) -\coth\left(\frac{\omega}{2T_{\mathrm{ph}}}\right)\right], \hspace{15pt} \Delta_0^3 = \hbar^2 \rho_{2d} u_{\mathrm{t}}^4
\end{align}
Performing the integration leads to the result in Eq.~\eqref{e-ph-cooling}.

\section*{Acknowledgments}
We gratefully acknowledge support from EPSRC under the grant EP/R029075/1. We thank V.\ I.\ Yudson for useful comments.

\end{document}